# The impact of software complexity on cost and quality – A comparative analysis between Open source and proprietary software

Anh Nguyen-Duc

*IDI, NTNU*

anhn@idi.ntnu.no

## Abstract

*Early prediction of software quality is important for better software planning and controlling. In early development phases, design complexity metrics are considered as useful indicators of software testing effort and some quality attributes. Although many studies investigate the relationship between design complexity and cost and quality, it is unclear what we have learned beyond the scope of individual studies. This paper presented a systematic review on the influence of software complexity metrics on quality attributes. We aggregated Spearman correlation coefficients from 59 different data sets from 57 primary studies by a tailored meta-analysis approach. We found that fault proneness and maintainability are most frequently investigated attributes. Chidamber & Kemerer metric suite is most frequently used but not all of them are good quality attribute indicators. Moreover, the impact of these metrics is not different in proprietary and open source projects. The result provides some implications for building quality model across project type.*

## Keywords

**Design Complexity, Software Engineering, Open source software, Systematic literature review**

## 1. Introduction

With the appearance of ultra-large software, system-of-system, Internet-of-Things, software became much larger and more complex that it was before. As software becomes more and more popular in not only computer, but also embedded systems, the users demand more reliable and powerful software. Software complexity has been shown to be one of the main contributing factors to the software development and maintenance effort [2]. Most experts today agree that complexity is a major feature of computers software, and will increasingly be in the future.

Software metrics, as a tool to measure software development progress, has become an integral part of software development as well as software research activities. Tom DeMarco stated that "You cannot control what you cannot measure" [1]. The large part of software measurements is to, in one way or another, measure or estimate software complexity due to its importance in practices and research. Measurement of software complexity enhances our knowledge about nature of software and indirectly assesses and predicts final quality of the product. It is claimed that complexity metrics is practically meaningful if they can indicate the projects outcomes, i.e, software quality and effort.

Literature in software engineering reveals a large number of proposed complexity metrics. However, it is not obvious to select appropriate metric set that can predict given software attributes. Which quality attributes are influenced by software complexity? Which metrics are useful for predicting these quality attributes? Knowing the proper metrics in a specific context would be beneficial for software industry. Moreover, it is interesting to know whether these metrics has the same effectiveness across different project context, for example open source and closed source projects. Since open source is emerging as an alternative to closed source development in software industry [25, 26], empirical comparison between open sources and proprietary software in multiple perspectives, such as software complexity, quality and development effort is always of highly interest.

These questions are non-trivial and can only be answered by proper empirical studies. Typically, such studies validate the usage of metrics by building a prediction model. The predictive power depends not only on the selection of metric suite but also prediction techniques and evaluation method. The available data and feature of the data set also influences the predictive result. While prediction techniques and evaluation method has been well-studied and systematically aggregated into a body of knowledge [3, 4, 5], there is very limited number of papers summarizing evidence-based knowledge about software complexity metrics [6]. Among the prediction models with the same prediction techniques, some studies yield different predictive results of same metric suite, even contradictory ones. This adds to the impression that, despite a large number of design metrics used in quality prediction models, it is currently still unclear whether we have learned anything from these studies. This paper presents a systematic review on the influence of software complexity metrics on quality attributes. We aggregated Spearman correlation coefficients from 59 different data sets from 57 primary studies by a tailored meta-analysis approach.

The rest of the paper is organized as follows. Section 2 presents a theory of cognitive complexity and statistical methods on software engineering. Section 3 states our research questions and research methodologies. While Section 4 presents the review process, Section 5 provides research results and discussion. Section 6 identifies the threats to validity. The paper ends with conclusions and future works.

## 2. BACKGROUND

### 2.1. Theory of cognitive complexity

IEEE standard defines software complexity as "*the degree to which a system or component has a design or implementation that is difficult to understand and verify*" [7]. The definition differentiates two different kind of complexity: complexity of design artifacts, such as UML diagram and complexity of implementation like source code. Basically, design complexity is structural features of design artifacts that can be measured prior to implementation stage. This information is important for predicting quality of final products in early phase of software development life cycle.

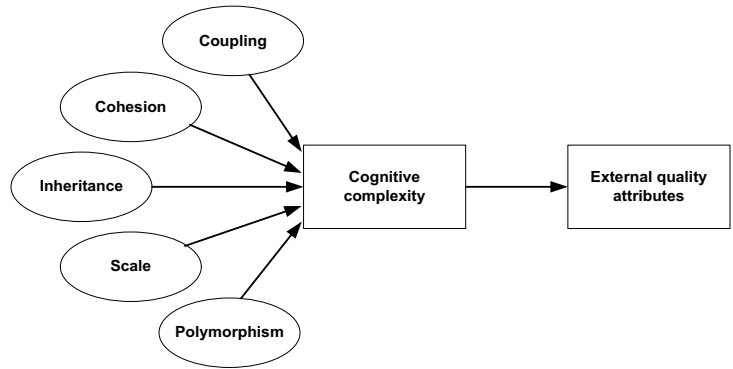

**Figure 1: Theory of cognitive complexity (adapted from [9])**

There is a theoretical basic for investigation the relation between design complexity and software external quality attributes, namely theory of cognitive complexity [8]. Briand et al. hypothesized that the structural properties of a software component have an impact on its cognitive complexity [8]. In his study, cognitive complexity is defined as "the mental burden of the individuals who have to deal with the component, for example, the designer, developers, testers and maintainers" [8]. Figure 1 presents a model of cognitive complexity theory, which stated that high cognitive complexity leads to the increasing effort to understand, implement and maintain the component [9]. As the results, it could lead to undesirable external qualities, such as increased fault-proneness and reduced maintainability.

**2.1. Empirical investigation on software complexity**

In the literature of research on software engineering, the emerging trend is explanatory empirical methods, which use empirical evidence to confirm or model the casual relationships between control variables and performance variables. The two common empirical investigation methods on software complexity are correlation analysis and statistical regression.

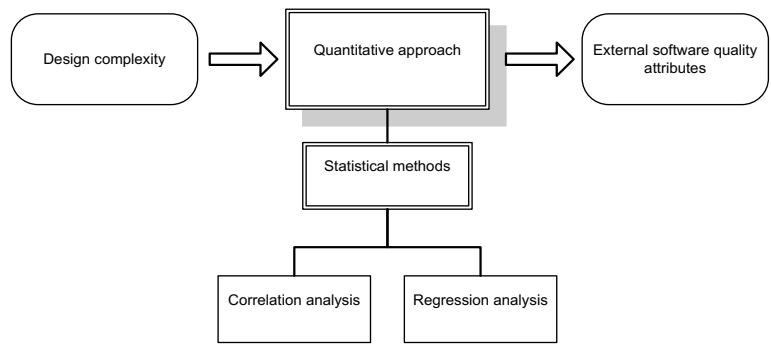

**Figure 2: Quantitative approaches for empirical validation in software engineering**

Correlation analysis investigates the extent to which changes in the value of an attribute (such as value of complexity metric in a class) are associated with changes in another attribute (such as number of defect in a class). An correlation coefficient represents a numerical summary of the degree of association between two variables, e.g., to what degree do high values of coupling of a class go with high number of defect in that class [27]. Though correlation between two variables does not necessarily result in causal effect [27], it is still an effective method to select candidate variables for causation relationship.

Regression goes beyond correlation by adding prediction capabilities. Regression analysis includes techniques for modeling and analyzing one or several variables to identify the relationship between a dependent variable and one or more independent variables [28]. Among large amount of regression techniques has been used in software engineering literature, the most common techniques are linear regression model and logistics regression model [28]. In both methods, it is necessary to control all confounding variables in order to achieve a meaningful casual relationship.

## 3. RESEARCH DESIGN

### 3.1. Research questions

Table 1 shows an adapted Goal-Question-Metric (GQM) template to form the research objective. The target objective of this research is software design complexity, in particular design complexity of OO system. The purpose is to externally validate design complexity by investigating its statistical relationship to external software quality. The topic is considered in the viewpoint of measurement researcher as well as practitioner. The study population is software engineering literature.

**Table 1: GQM template of research goals**

| | |
|---|---|
| Analyze | Software design complexity |
| For the purpose of | Confirming |
| With respect to their | Statistical relationship to cost and external software quality. |
| From the view point of | Measurement practitioner, researcher |
| In the context of | Software engineering literature |

The main research objective is refined as four research questions:

- RQ1: Which quality attributes are influenced by design complexity metric in literature?
- RQ2: What are the most investigated design complexity metrics in literature?
- RQ3: Which design complexity metrics are potentially predictors of quality attribute?
- RQ4: How is the influence of design complexity on software quality in open source vs. closed source software?

### 3.2. Research methodology

Figure 3 presents research methods used to answer the stated questions.

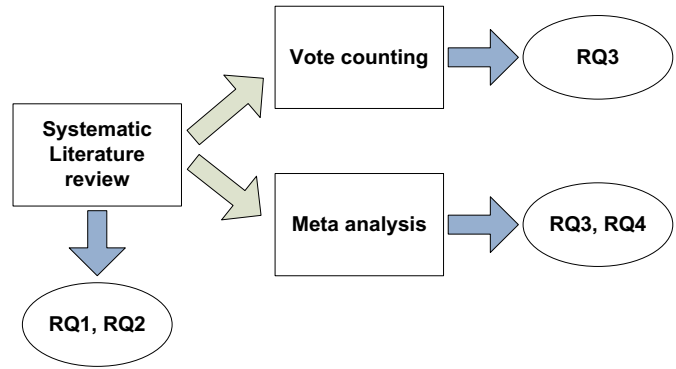

**Figure 3: Research methodology**

### 3.2.1. Systematic literature review

A systematic literature review (often referred to as a systematic review) is a common method for identifying, evaluating and interpreting all available research relevant to a particular research question, or topic area, or phenomenon of interest. This method is very common in research of medicine and social science [10]. Individual studies contributing to a systematic review are called *primary* studies; a systematic review is a form of *secondary* study [10]. The most common reasons are to summarize the existing knowledge about interested research questions, to identify any gaps in current research in order to suggest areas for further investigation and to provide a framework/background in order to appropriately position new research activities. The systematic review involves the following steps, with the detail description provided in [10]:

- Specifying research questions
- Data source selection
- Search strategy development
- Search string formation
- Study selection criteria identification
- Study quality assessment identification
- Study extraction strategy identification

There are two approaches to quantitatively summary results from systematic review, namely vote counting and meta analysis [11].

### 3.2.2. Vote counting

Vote counting is an alternative of meta analysis when there is not enough of statistical reported data. The main advantage of vote counting is that it is conceptually simple and can be used with very little information [11]. Besides it does not require actual value of effect sizes for aggregation [11]. The only assumption of the method is that studies use the same kind of statistical test. Vote counting, however, is not free from error. Vote counting can be strongly biased towards the conclusion of zero effect size in populations of studies with small effect size and a low number of subjects. Besides, this bias is not reduced as the number of studies increase [12]. Therefore, the result of vote counting should be confirmed by other statistical methods, such as meta analysis.

### 3.2.3. Meta analysis

Meta-analysis is a statistical technique for identifying, summarizing, and reviewing results of quantitative studies to arrive at conclusions about a body of research [13, 14]. A typical meta-analysis is to summarize all the research results on one topic and to discuss reliability of the summary. It is noticed that meta-analysis is only recently introduced in software engineering [11, 15, 16]. Meta analysis is selected because of:

- Its well-founded mathematic background and statistical validity
- Its ability to provide a numerical aggregation value of studies. The outcome of a meta-analysis is an average effect size with an indication of how variable that effect size is between studies.
- Its ability to deal with study conflict.
- Its ability to deal with publication bias.

## 4. RESEARCH CONDUCT

Table 2 shows the search protocol. We choose Scopus, IEEE Explore and ACM Digital Library as search databases due to its functionality, reputation and familiarity to the reviewer. As common search fields in systematic reviews [3, 4, 5] article title, keyword and abstract will be used in this work. With the objective to retrieve all relevant studies, we use the wide search range among research on Software and System Engineering. The only limitation is that we only search in publication written in English.

**Table 2: Search protocol**

| Search database | Scopus, IEEE Explore, ACM Digital Library |
|---|---|
| Search field | Title, Abstract, Keyword |
| Search topic | Software and system engineering |
| Language | English |
| Search string formula | *("software complexity" Or Synonyms) AND("impact" Or Synonyms) AND ("cost" or "quality" or Synonyms) AND software* |

The search string query from Scopus database resulted in a total of 906 primary publications. We excluded 625 papers out of a total of 906 primarily based on title and abstract, which were clearly out of scope and did not relate to any research questions. The remaining 281 papers were subject to detailed exclusion criteria. The paper is scanned to find out whether any exclusion criteria are met. This stage resulted in remaining 85 papers, which were further filtered out by reading full text. The studies are carefully read to find whether paper's objective, study design, data collection and analysis are relevant to answer our questions. This step remains 31 papers. The second search in IEEE Explore and ACM Digital Library resulted in a total of 265 papers. It is noticed that many papers in this set is duplicated with previous pool because Scopus also index papers in IEEE Explore and ACM Digital Library. After eliminating duplicated papers, there were left 85 papers. The title and abstraction exclusion results in 40 remaining papers and further exclusion based on detailed criteria reject 25 papers out of this 40. The remaining 15 papers are read and finally 8 of them are selected.

It is revealed that no relevant papers are found prior to 1996. It was also the time when the Empirical Software Engineering journal was firstly published. It is worth noticed that in the late 90's, there were not so much publications and the most probable reason for this is that design metric was a new research area at that time, i.e. Chidamber & Kemerer metric suite was firstly introduced in 1991 and refined in 1996. The figure also shows the increasing number of studies between 5-year periods. From 1995 to 1999, there are 9 relevant papers while from 2000 to 2004; the numbers of relevant papers are 15. This number is 33 for the period of 5 years from 2005 to 2009, which is more than double the number of previous period. This fact indicates the increasing interest of research community in the topic recently.

## 5. RESULT AND DICUSSION

### 5.1. RQ1: Which quality attributes are influenced by design complexity metric in literature?

**a. Results**

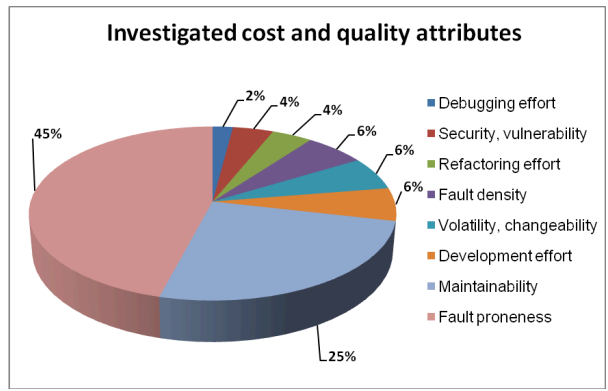

**Figure 4: Investigated quality attributes**

Figure 4 shows the external quality attributes that are investigated for the relationship with design complexity. There are four main investigated costs and quality attributes, namely reliability, maintainability, reusability and development effort that are refined in 9 sub categories as below:

- Defect density: includes studies that investigate relation between total known defects divided by the size (defect density) and design complexity.
- Fault proneness: includes studies that investigate the relation between probabilities of finding a defect in a class (fault proneness) with design complexity.
- Vulnerability, security: includes studies that investigate the relation between numbers of faults that lead to security failure and design complexity
- Testability: includes studies that predict effort required for software testing due to level of design complexity [17]
- Maintainability: includes studies that predict time, efforts required to correct bugs, improve performance or adapt software to a change environment [7].
- Volatility: studies that investigate the relation between the numbers of enhancement in a class with design complexity.
- Debugging cost: includes studies that investigate the relation between effort and time for fixing a class with its design complexity.
- Development effort: include studies that investigate the relation between cost and time for development with its design complexity.
- Refactoring effort: include studies that investigate the relation between efforts for refactoring a class with its design complexity.

**b. Discussion:**

The statistics description shows the most investigated cost and quality attributes are fault proneness and maintainability. Since the information about design complexity is theoretically available in the design phase, many primary studies collect these data in the implementation or maintenance phase. Hence, the post-design phase software attributes is more popular subject of research. Among reliability related attributes, fault-proneness is investigated much more than fault density or security. The reason could come from the fact that the data in number of fault per software unit is not sufficient. Therefore, the quality model is more on the level that predict whether a software unit is faulty or not. Due to the requirements of dataset size for aggregation, only maintainability and fault proneness studies have sufficient combinable data. Hence, from next questions, we only focus on these two attributes.

**5.2. RQ2: What are the most investigated design complexity metrics in literature?**

**a. Result:**

Table 3 shows the list of metrics, which are used in more than 50% of total selected primary studies. The most frequent used metrics belong to adjusted version of Chidamber & Kemerer metric suite, namely, NOC, DIT, CBO, LCOM version 2, WMC and RFC version 1. Among 120 investigated metrics, we found that the distribution of usage of design metrics are followed Pareto-like distribution 20/60: the 20% most common metrics are investigated in 60% of the studies.

**Table 3: List of frequent used metrics**

| Metric | Definition | Type |
| --- | --- | --- |
| NOC | number of classes that directly inherit from a given class | inheritance |
| DIT | length of the longest path from the class to the root of inheritance hierarchy [18] | inheritance |
| CBO | number of other classes to which it is coupled, included inheritance-based coupling [18] | coupling |
| LCOM2 | LCOM1 minus count number of pairs of methods that do (average percentage of the method in a class using an attribute and subtract from 100%) [18] | cohesion |
| WMC | number of all methods of a class [18] | size |
| RFC1 | RFCinf + not count methods indirectly invoked by methods in M [18] | coupling |

**b. Discussion:** The domination of Chidamber & Kemerer metric set in correlation and prediction studies could comes from the fact that they are one of the first complete design metrics with measure of all OO features [18]. The metric set has strong foundation since they were developed based on the ontology of Bunge and validated against Weyuker's measurement theory principles [18]. More importantly, Chidamber et al. claimed that these measures can aid users in understanding design complexity, in detecting design flaws and in predicting certain project outcomes and external software qualities such as software defects, testing, and maintenance effort and provided an empirical sample of these metrics from two commercial systems [8]. Besides, the metric set is simple, well-understood [19] and straightforward to collect.

These reasons seem to invoke great enthusiasm among researchers and software engineers, and a great amount of empirical studies have been conducted to evaluate those metrics. Over time, this metric suite is gradually growing in industry acceptance. In particular, they are being incorporated into development tools such as Rational Rose [20] and JBuilder [21]. Many others tool such as WebMetric [23] and JMetric [22], also auto collection feature for the metric set and make it very easy to collect.

### 5.3. RQ3: Which design complexity metrics are potentially predictors of quality attribute?

To find whether a design metric is a potential predictor of external attributes, we test each design metric with the following hypothesis:

*H0: There is no impact of metric X on quality attribute Y.*

Due to the variation of the use of design metrics in primary studies, we use vote counting to test the hypothesis H0. Assumed that each study report a significant value (p value) of correlation coefficient or regression coefficient between design metric X and quality attribute Y. It is worth noticed that some studies report the actual value of the significance test, but some others only report the significant level, which is represented by + if the correlation test is significant at 0.05 and by ++ if the test is significant at 0.01. To maximize number of included studies, we select the significance level at 0.05. The null hypothesis is rejected if the portion of success in total studies is greater than 0.5. Otherwise, we accepted the hypothesis.

Among 120 design metrics investigated in fault proneness studies and 59 design metrics investigated in maintainability studies, we select the metrics whose significance level are reported in at least 6 different datasets.

**a. Result for Fault proneness**

Table 4 describes the null hypothesis test for significant positive impact of several design metrics on fault proneness in Spearman correlation result. Table 4 presents number of studies, total number of dataset, and number of datasets that a metric has significantly positive and negative correlation coefficient, number of datasets that a metric is non-significant.

There are 12 design metrics that are investigated in more than 6 data sets. Out of these 12 metrics, 9 metrics has portion of success more than 50%, namely CBO, WMC, RFC, WMC Mc Cabe, SDMC, AMC, NIM, NCM and NTM and therefore null hypothesis is rejected for these metrics. The 3 other metrics that have less than 50% of success portion are NOC, DIT, LCOM. Hence, we accept the null hypothesis: with spearman test, there is no positive impact of NOC, DIT or LCOM on fault proneness.

**Table 4: Hypothesis test for significantly Spearman coefficient for fault proneness**

| Metric | No of studies | No of data set | No of + | No of - | No of non significant | % of + | Positive hyp. test |
|---|---|---|---|---|---|---|---|
| NOC | 8 | 19 | 6 | 1 | 12 | 32% | accept |
| DIT | 7 | 14 | 2 | 0 | 12 | 14% | accept |
| CBO | 8 | 17 | 10 | 0 | 7 | 59% | reject |
| LCOM | 7 | 14 | 6 | 0 | 8 | 43% | accept |
| WMC | 9 | 26 | 18 | 0 | 8 | 69% | reject |
| RFC | 7 | 15 | 9 | 0 | 6 | 60% | reject |
| WMC Cabe | 3 | 16 | 11 | 0 | 5 | 69% | reject |
| SDMC | 1 | 6 | 6 | 0 | 0 | 100% | reject |
| AMC | 1 | 6 | 6 | 0 | 0 | 100% | reject |
| NIM | 1 | 6 | 6 | 0 | 0 | 100% | reject |
| NCM | 1 | 6 | 6 | 0 | 0 | 100% | reject |
| NTM | 1 | 6 | 6 | 0 | 0 | 100% | reject |

**b. Result for Maintainability**

Table 5 describes the null hypothesis test in Spearman correlation in maintainability studies. There are 6 design metrics that are investigated in more than 6 data sets. Out of these metrics, 4 metrics has portion of success more than 50 percent and therefore we can reject null hypothesis in these metrics. The other 2 metrics that have less than 50% of success portion are NOC and DIT. Therefore, we accept the null hypothesis: with spearman test, there is no positive impact of NOC, DIT or LCOM on fault proneness. The complete list of metrics is given in [29].

**Table 5: Hypothesis test for significantly Spearman coefficients for maintainability**

| Metric | No of studies | No of data set | No of + | No of - | No of non significant | % of + | Positive hyp. test |
|---|---|---|---|---|---|---|---|
| WMC | 5 | 7 | 6 | 0 | 1 | 86 | reject |
| RFC | 6 | 8 | 8 | 0 | 0 | 100 | reject |
| DIT | 3 | 5 | 2 | 0 | 3 | 40 | accept |
| NOC | 4 | 6 | 2 | 0 | 4 | 33 | accept |
| MPC | 3 | 6 | 6 | 0 | 0 | 100 | reject |
| DAC | 5 | 6 | 4 | 0 | 2 | 67 | reject |

**c. Discussion:**

For both fault proneness and maintainability studies, vote counting shows that there is no evidence about relationship of NOC, DIT to external quality attributes. These metrics measure the inheritance dimension of object oriented systems. WMC, RFC is shown to be potential good predictors for external quality attributes in all the cases. In fault proneness studies, NIM, NCM, NTM are scale-based metrics when CBO, SDMC, AMC are coupling-based metrics. In Maintainability studies, DAC and MPC are also coupling-based metrics. This observation adds an impression that design metrics that capture scale and coupling dimensions are useful in indicating the final product quality attributes, such as fault proneness and maintainability. The result also indicates that, it is unlikely inheritance and cohesion-based metrics can show significant relationship with software quality attributes.

In comparison with the most frequent use design metrics, the results show that for fault-proneness and maintainability study, only 50% of frequently used metrics are potentially good ones. Alternatively, the common metrics are not necessarily the good ones. This suggests that the researchers in quality model should evaluate the metric effectiveness before adding them to the model.

### 5.4. RQ4: How is the influence of design complexity on software quality in open source vs. closed source software?

**a. Result:**

There are 35 private datasets, which contributes 59% of total number of datasets. Among them, 10 datasets come from academic projects and 24 datasets come from industry project. The remaining 24 dataset are public, which contributes 41% of total number of datasets. Figure 6 illustrate the aggregated results for the correlation between metric WMC and Fault proneness using Meta analysis. The box center represent for the correlation coefficient, the box size represents for the weight of the dataset, using the dataset size. The line shows the 95% confidence interval of the correlation coefficient.

Table 6: Correlation intervals in closed source vs. Open source

| Metric | Closed source | | Open source | | % diff. |
|---|---|---|---|---|---|
| | 95% | # | 95% | # | |
| NOC | [-0,18;-0,03] | 9 | [0,02; 0,05] | 9 | 24% |
| DIT | [0,08; 0,23] | 13 | [0,04; 0,07] | 9 | 0.2% |
| CBO | [0,16; 0,31] | 15 | [0,26; 0,30] | 9 | 4.3% |
| LCOM | [0,17; 0,32] | 9 | [0,15; 0,18] | 9 | 0.0% |
| RFC | [0,22; 0,37] | 8 | [0,24; 0,27] | 9 | 2.9% |
| WMC | [0,20; 0,35] | 15 | [0,27; 0,29] | 18 | 3.9% |

Table 5 shows the influence of different metrics on fault proneness across project type. The table reports the 95% confidence interval of aggregated coefficients for metrics across project type and the number of investigated datasets. Besides, the difference between the two intervals is calculated by variance explanation of the separator variable [13, 14]. From the statistics, we have an observation:

- The correlations between design metrics and fault proneness are at trivial to medium level.
- The Project type variable does not cluster the correlation of metrics and fault proneness since the explanation power is very small (0 to 4.3%). Only with NOC, the variance explanation is 24% with two non-overlaps confidence intervals.
- The 95% confidence intervals of metric correlation in closed source projects are larger in those in open source projects.

**b. Discussion:**

The small variance explanation shows that project type is not a good separator variable. Alternatively, there are no difference in relationship between design metrics and fault proneness between closed source and open source projects. However, the confidence intervals in open source projects are smaller, which indicate a more homogeneous collection of datasets than closed source datasets group.

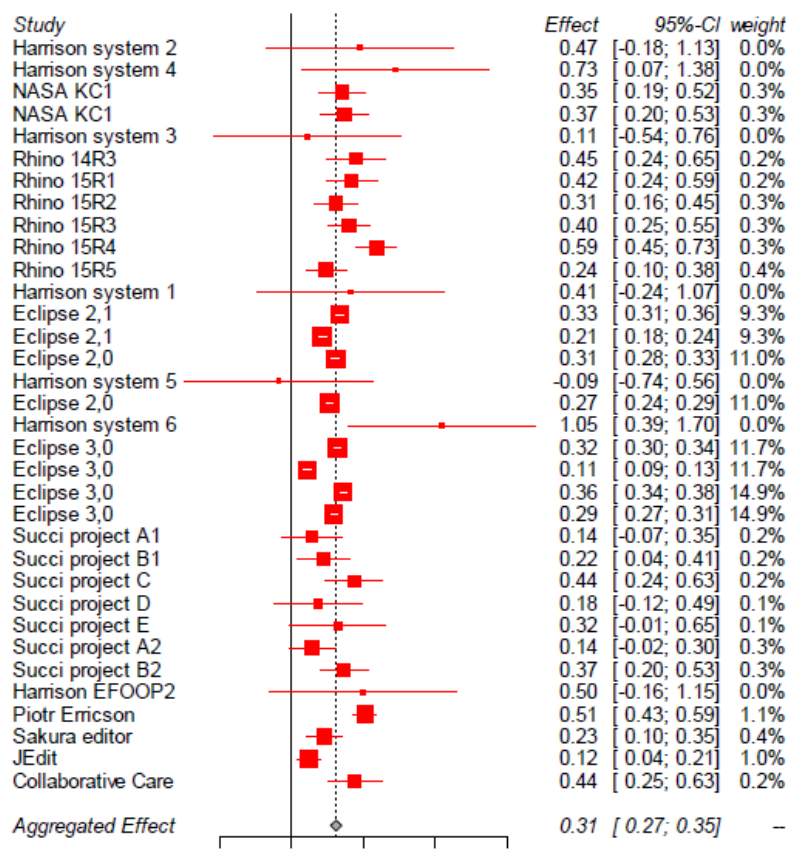

**Figure 5: Aggregated correlation coefficient of WMC in fault proneness studies**

## 6. THREATS TO VALIDITY

The major threats to the study validity come from the systematic review and data aggregation process, which are quality of search process, univariate analysis and aggregation of heterogeneous data.

The first validity threat comes from the search quality. We use two measures to evaluate the search result, namely recall and precision [24]. Since it is impossible to know all existing relevant items, the recall of the search string was estimated by conducting a pilot search, which showed an initial recall of 88%, and after a refinement of the search string, a recall of 100%. Besides, it was not tried to optimize the search string for precision, which is clearly reflected by the final, very low, precision value of 6.3% (considering 906 documents after the removal of duplicates and 57 selected primary studies).

Another main limitation of our study is that it focuses on relationships between single design metric and quality attribute, such as fault proneness and maintainability. However, it may be that a combination of design metrics leads to a better prediction than single ones. Therefore, it may be that multivariate models that use several design metrics to predict fault-proneness are more meaningful for this purpose. However, the existing multivariate models are so heterogeneous that it may be impossible to generalize from them.

Last but not least, the primary studies used in this paper stem from various sources with different quality. One typical critique of meta-analysis is that combining such studies equals mixing apples and oranges. It can be argued, however, that mixing apples and oranges is beneficial, particularly if one wants to generalize about fruit, and that studies that are exactly the same in all respects are actually limited in generalizability [10].

## 3. CONCLUSIONS

Table 7 summarizes the results for the research questions.

Table 7: Summary of findings

| Question | Answers |
|---|---|
| RQ1: Which quality attributes are predicted using design complexity metrics? | Reliability, maintainability, reusability and development effort |
| RQ2: Which design complexity metrics is most frequently used in literature? | NOC, DIT, CBO, LCOM2, WMC and RFC1 |
| RQ3: Which design complexity metrics are potentially predictors of quality attribute? | Fault proneness: CBO, WMC, RFC, SDMC, AMC, NIM, NCM and NTM. Maintainability: WMC, RFC, MPC and DAC |
| RQ4: How is the influence of design complexity on software quality in open source vs. closed source software? | Larger variation in closed source projects. No difference in effectiveness toward fault proneness |

Our findings contribute to both research community and software practitioner in many ways. To researchers who are interested in software quality model, we confirm the importance of software coupling and scale metrics. These two dimensions should be more focused than inheritance and cohesion in fault prediction models. We suggest a list of specific design metrics that are potential predictors of fault proneness and maintainability. We also find a limited number of studies focus on other cost and quality attributes, such as development effort or fault density. These cost and quality attributes would be a research gap for further studies. The results confirm that, regarding to the impact of design complexity on quality is heterogeneous across project types. To software practitioners, since there is no separated regions found for open source and closed source projects, the quality prediction models on open source projects could be tried on closed source environment. This is helpful for industrial practitioners since they can learn from the quality model that is published using open source datasets.

The results of this thesis can be considered as a cornerstone for further study in the area of correlation and regression analysis of design complexity. The future works could focus on aggregating the multivariate quality models and investigating other separator variables than project type. Besides, the cost and quality attributes such as security, volatility or refactoring effort are potential targets for future investigation of software complexity.

## Authors

Dr. Anh Nguyen-Duc is a researcher in the Norwegian University of Science and Technology's Department of Computer and Information Science. His research focuses on Empirical Software Engineering, Global Software Development, Socio-technical system and Software Startups

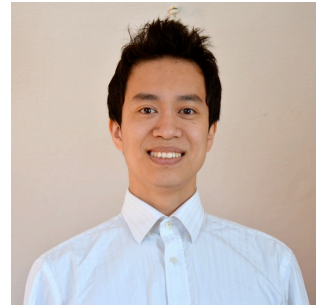